\documentclass[twocolumn,showpacs,amsmath,amssymb,pra]{revtex4}

\usepackage{bm}
\usepackage{color}
\usepackage{graphicx}   
\newcommand{\re}{{\mathrm e}}
\newcommand{\ri}{{\mathrm i}}

\begin{document}

\title[Fluctuations of order parameter]
      {Fluctuations of the order parameter of a mesoscopic Floquet condensate}

\author{Bettina Gertjerenken}
\email{b.gertjerenken@uni-oldenburg.de}

\author{Martin Holthaus}

\affiliation{Institut f\"ur Physik, Carl von Ossietzky Universit\"at, 
	D-26111 Oldenburg, Germany}
                  
\date{October 23, 2014}

\begin{abstract}
We suggest that nonequilibrium Bose-Einstein condensates may occur in 
time-periodically driven interacting Bose gases. Employing the model of 
a periodically forced bosonic Josephson junction, we demonstrate that 
resonance-induced ground state-like many-particle Floquet states possess 
an almost perfect degree of coherence, as corresponding to a mesoscopically 
occupied, explicitly time-dependent single-particle orbital. In marked 
contrast to the customary time-independent Bose-Einstein condensates, the 
order parameter of such systems is destroyed by violent fluctuations when 
the particle number becomes too large, signaling the non-existence of a 
proper mean field limit. 
\end{abstract} 



\maketitle


\section{Nonequilibrium condensates}

In the wake of traditional textbook teaching, Bose-Einstein condensation 
usually is associated with thermal equilibrium: At sufficiently low
temperatures a Bose gas ``condenses'' into the lowest single-particle 
state~\cite{LandauLifshitz75,Huang87,PathriaBeale11}. In the present paper 
we take a theoretical step towards the exploration of nonequilibrium 
condensates~\cite{VorbergEtAl13}.

The possible existence of such nonequilibrium condensates is reflected in
the fundamental Penrose-Onsager criterion~\cite{PenroseOnsager56} for 
Bose-Einstein condensation in a system of $N$ repulsively interacting Bose 
particles, where $N$ is large: This criterion does neither require thermal 
equilibrium nor even steady states~\cite{Leggett01}. Instead, it takes 
recourse to the one-particle reduced density matrix
\begin{equation}
	\varrho(\bm r, \bm r';t) = \langle \Psi_N(t) |
	\widehat{\psi}^\dagger(\bm r) \widehat{\psi}(\bm r') | 
	\Psi_N(t) \rangle \; ,
\label{eq:OPR}
\end{equation}  
where $| \Psi_N(t) \rangle$ denotes the state of the $N$-Boson system at
time $t$, and $\widehat{\psi}^\dagger(\bm r)$ and $\widehat{\psi}(\bm r)$ are 
the usual creation and annihilation operators, obeing the Bose commutation 
relation $\left[ \widehat{\psi}(\bm r), \widehat{\psi}^\dagger(\bm r') \right]
= \delta(\bm r - \bm r')$. Considered as a matrix with indices $\bm r$ and
$\bm r'$, its diagonal elements $\varrho(\bm r, \bm r;t)$ provide the particle 
density of the system at the position $\bm r$. Because at each moment this 
matrix is Hermitian, it can be decomposed in terms of a complete set of 
orthonormal single-particle functions $\chi_j^{\phantom *}(\bm r;t)$ with 
eigenvalues $n_j(t)$, such that
\begin{equation}
	\varrho(\bm r, \bm r';t) = \sum_j n_j(t)
	\chi_j^{\phantom *}(\bm r,t) \chi_j^*(\bm r', t) \; .	
\label{eq:SDD}
\end{equation} 
According to Penrose and Onsager one has a simple Bose-Einstein condensate
when the largest eigenvalue $n_{\max}(t)$ is on the order of $N$, all others
being of order~$1$; the corresponding eigenfunction $\chi_{\max}(\bm r;t)$ 
then is the condensate wave function~\cite{PenroseOnsager56}. In the most 
favorable case where $n_{\max}(t) = N$, the density matrix~(\ref{eq:SDD}) 
reduces to a projector, times $N$, onto the $N$-fold occupied single-particle 
orbital $\chi_{\max}(\bm r;t)$. As a matter of principle, this orbital can 
have an arbitrarily strong time-dependence.

Here we suggest that a particular type of nonequilibrium condensate may 
become experimentally accessible when an interacting Bose gas is subjected 
to a resonant time-periodic force. In general, when a quantum system evolves 
according to a Hamiltonian $H(t) = H(t+T)$ which depends periodically on 
time with period~$T$, and remains bounded, the Floquet theorem asserts that 
there exists a complete set of solutions to the time-dependent Schr\"odinger 
equation which possess the particular form 
$|\psi_j(t)\rangle = | u_j(t) \rangle \exp(-\ri\varepsilon_j t/\hbar)$, 
where the Floquet functions $| u_j(t) \rangle =  | u_j(t+T) \rangle$ inherit 
the imposed periodicity in time, and the quantities $\varepsilon_j$ which 
determine the growth rates of the accompanying phases are known as 
quasienergies~\cite{Shirley65,Zeldovich67,Sambe73,FainshteinEtAl78}. Each 
solution to the time-dependent Schr\"odinger equation can be expanded in 
this Floquet-state basis with constant coefficients, implying that one can 
describe, e.g., a time-periodically driven ideal Bose gas by means of 
single-particle Floquet orbitals which carry constant occupation 
numbers~\cite{VorbergEtAl13}. In particular, it makes sense to introduce 
the notion of a macroscopically occupied Floquet state. 

Recent experiments with Bose-Einstein condensates in optical lattices 
subjected to strong time-periodic forcing already have demonstrated dynamic 
localization~\cite{LignierEtAl07,EckardtEtAl09,ArimondoEtAl12}, coherent
control of the superfluid-to-Mott insulator transition~\cite{ZenesiniEtAl09},
giant Bloch oscillations~\cite{AlbertiEtAl09,HallerEtAl10}, 
frustrated classical magnetism~\cite{StruckEtAl11}, 
controlled correlated tunneling~\cite{ChenEtAl11}, 
artificial tunable gauge fields~\cite{StruckEtAl12,StruckEtAl13},
and effective ferromagnetic domains~\cite{ParkerEtAl13}.
Without claiming completeness of this list, these experiments testify that a 
macroscopic matter wave persists in the presence of strong time-periodic 
forcing.

\section{Appearance of new ground state}

For our theoretical considerations we employ the model of a 
periodically driven bosonic Josephson junction, which can be realized, 
for instance, with Bose-Einstein condensates in optical double-well 
potentials~\cite{GatiMKO07}. The junction itself is described by the 
Lipkin-Meshkov-Glick Hamiltonian~\cite{LipkinEtAl65}
\begin{equation}
	H_0 = -\frac{\hbar\Omega}{2} 
	\left(a^{\phantom{\dagger}}_1 a_2^{\dagger} 
	+ a_1^{\dagger}a^{\phantom{\dagger}}_2\right) 
	+ \hbar \kappa \left(
	a_1^{\dagger} a_1^{\dagger} 
	a^{\phantom{\dagger}}_1 a^{\phantom{\dagger}}_1 
	+ a_2^{\dagger} a_2^{\dagger}
	a^{\phantom{\dagger}}_2 a^{\phantom{\dagger}}_2 \right) \; ,
\label{eq:UDJ}
\end{equation}
where the operators $a_j^{\dagger}$ and $a^{\phantom{\dagger}}_j$ create and 
annihilate, respectively, a Bose particle in the $j$th well ($j=1,2)$, obeying 
the commutation relation 
$\left[ a^{\phantom{\dagger}}_j, a_k^{\dagger} \right] = \delta_{jk}$.
Moreover, $\hbar\Omega$ is the single-particle tunneling splitting, and
$2\hbar\kappa$ quantifies the repulsion energy of each pair of bosons occupying 
the same well. This Hamiltonian~(\ref{eq:UDJ}) had originally been devised for 
testing many-body approximation schemes~\cite{LipkinEtAl65}; its paradigmatic 
importance as a nontrivial, but well tractable model for interacting Bose gases
has been realized shortly after experiments with ultracold atomic vapors became
standard practice~\cite{MilburnEtAl97,ParkinsWalls98}. We extend this model
by assuming that the two wells are time-periodically shifted with frequency
$\omega$ in phase opposition to each other, giving rise to the total
Hamiltonian~\cite{HolthausStenholm01,JinasunderaEtAl06}    
\begin{equation}   
	H(t) = H_0 + \hbar\mu_1 \cos(\omega t) 
	\left( a_1^{\dagger} a^{\phantom{\dagger}}_1 
	- a_2^{\dagger} a^{\phantom{\dagger}}_2 \right) \; .
\label{eq:DJJ}
\end{equation}
Here the driving amplitude $\hbar\mu_1$ denotes the maximum shift in energy;
bosonic Josephson junctions with different driving schemes have also been 
considered in the literature~\cite{MahmudEtAl05,BoukobzaEtAl10}.

\begin{figure}[t]
\begin{center}
\includegraphics[width = 0.65\linewidth]{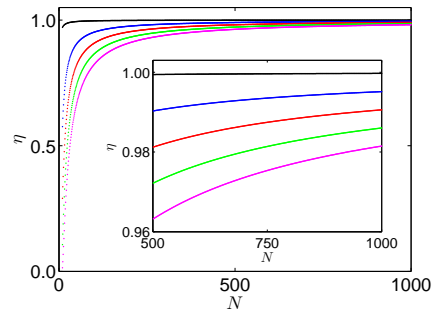}
\end{center}
\caption{(Color online) Degree of coherence~(\ref{eq:ETA}) for the lowest five 
	energy eigenstates $| j \rangle$ (top to bottom: $j = 0,\ldots, 4$) of 
	the undriven bosonic Josephson junction~(\ref{eq:UDJ}) with fixed 
	scaled interaction strength $N\kappa/\Omega = 0.95$ vs.\ particle 
	number~$N$.}   
\label{F_1}
\end{figure}

With the spatial degree of freedom being restricted to two discrete sites, the 
one-particle reduced density matrix~(\ref{eq:OPR}) becomes the $2\times 2$ 
matrix
\begin{equation}
	\varrho = \left( \begin{array}{cc}
	\langle a_1^{\dagger} a_1^{\phantom{\dagger}} \rangle &
	\langle a_1^{\dagger} a_2^{\phantom{\dagger}} \rangle \\
	\langle a_2^{\dagger} a_1^{\phantom{\dagger}} \rangle &
	\langle a_2^{\dagger} a_2^{\phantom{\dagger}} \rangle 
		\end{array} \right) \; ,
\label{eq:TBT}
\end{equation}		
where the expectation values are taken with respect to the state under 
consideration. If the junction is filled with $N$ particles, the 
Penrose-Onsager criterion now always confirms the existence of a condensate, 
but the question is whether this condensate is simple or fragmented: In the
former case the larger eigenvalue of the matrix~(\ref{eq:TBT}) is close to 
$N$, while the smaller is close to zero, thus indicating that there exists 
one single-particle state which is almost $N$-fold occupied. In contrast, the
condensate is fragmented when both eigenvalues are close to $N/2$. Therefore,
Leggett has introduced the quantity~\cite{Leggett01}    
\begin{equation}
	\eta  =2 N^{-2} \, {\rm tr} \, \varrho^2 - 1 \, ,
\label{eq:ETA}
\end{equation}	
computed from the trace of the squared density matrix, as an invariant measure 
of the degree of the system's coherence: One has $\eta = 1$ for a pure simple
condensate, whereas $\eta = 0$ in the case of maximum fragmentation. In 
Fig.~\ref{F_1} we plot $\eta$ for the lowest five energy eigenstates of the
undriven junction~(\ref{eq:UDJ}). Here the scaled interaction strength
$N\kappa/\Omega = 0.95$ is kept fixed as the particle number $N$ is varied, as
is required for approaching the mean field limit: In a rigorous mathematical
setting, that limit, which is described by the Gross-Pitaevskii theory,
requires $N \to \infty$ such that the product of particle number and 
interaction strength remains constant~\cite{LiebEtAl00}. Evidently, the
ground state $| 0 \rangle$ is almost fully coherent when $N$ becomes
sufficiently large, thereby indicating the existence of a {\em bona fide\/}
order parameter, namely, of a single-particle orbital which is occupied by
almost all of the $N$ particles when the system~(\ref{eq:UDJ}) is in its
ground state, and which thus constitutes the macroscopic wave function. It 
is well known that the exact ground state of the Hamiltonian~(\ref{eq:UDJ}) 
coincides with an exact coherent state only when $\hbar\kappa = 0$. However, 
the difference between the exact ground state $| 0 \rangle$ and an exactly 
coherent state here becomes insignificant when approaching the mean field 
limit, when $\hbar\kappa$ vanishes proportionally to $1/N$.

\begin{figure}[t]
\begin{center}
\includegraphics[width = 0.75\linewidth]{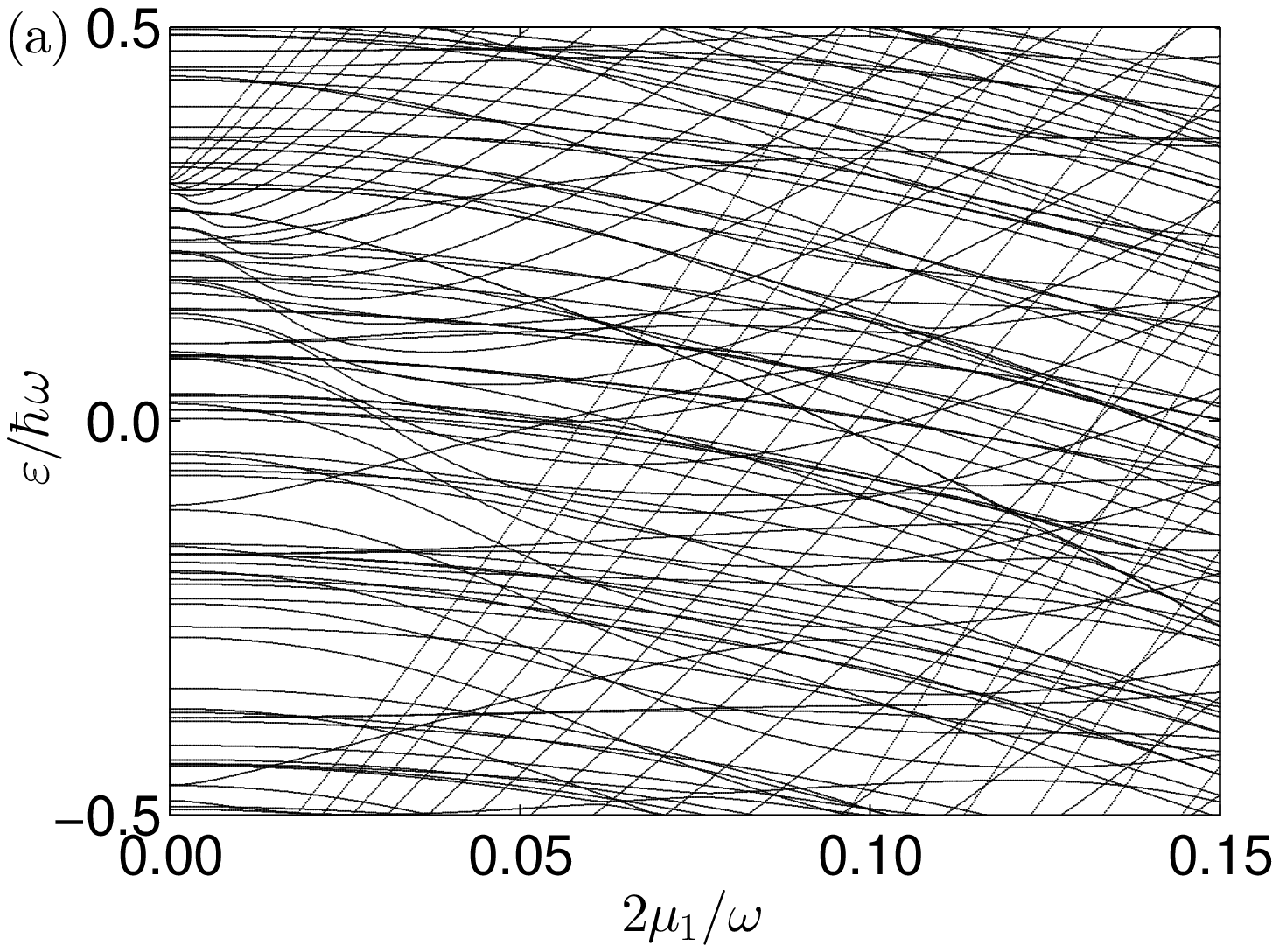}
\includegraphics[width = 0.75\linewidth]{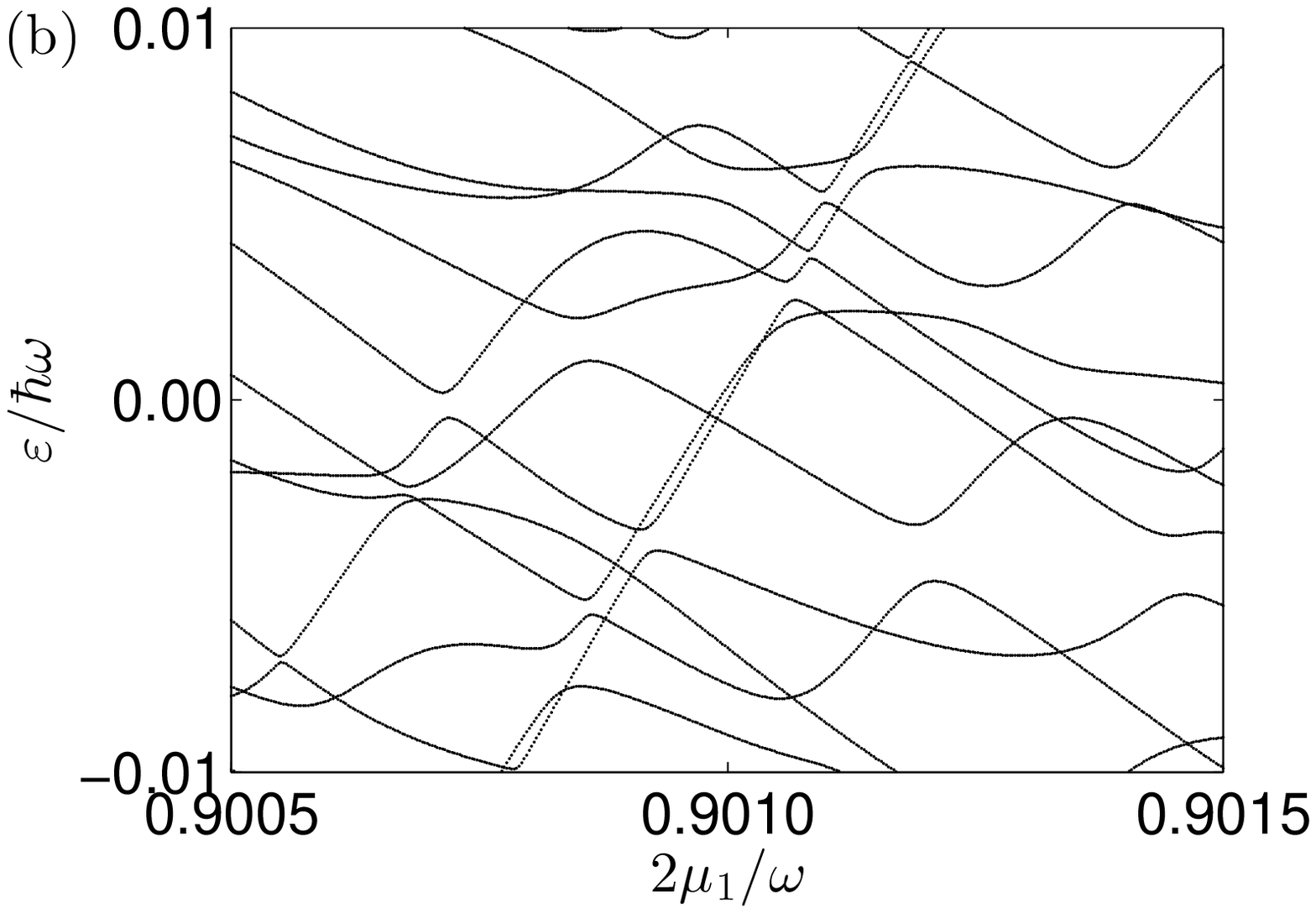}
\end{center}
\caption{(a) One Brillouin zone of exact quasienergies for the driven bosonic 
	Josephson junction~(\ref{eq:DJJ}) with $N = 100$ particles, scaled 
	interaction strength $N\kappa/\Omega = 0.95$, and scaled driving 
	frequency $\omega/\Omega = 1.62$, for low scaled driving amplitudes
	$2\mu_1/\omega$. The fan of almost equidistant lines is well described 
	by the Mathieu approximation~(\ref{eq:MAP}).
	(b) Part of the quasienergy spectrum for $N = 500$, and higher driving 
	amplitudes. Observe the scales!}   
\label{F_2}
\end{figure}

We now extend this analysis to the driven junction~(\ref{eq:DJJ}). Here we 
focus on {\em resonant\/} driving, {\em i.e.\/}, we choose the frequency 
$\omega$ such that $\hbar\omega$ equals the spacing $E_{r+1} - E_r$ of the 
unperturbed energy eigenvalues $E_j$ of the junction~(\ref{eq:UDJ}) at a 
particular state label $j = r$. Figure~\ref{F_2}~(a) shows the exact
quasienergies of the system for $N = 100$ particles, scaled interaction 
strength $N\kappa/\Omega = 0.95$, and scaled driving frequency 
$\omega/\Omega = 1.62$. This implies $r = 8$, so that the unperturbed 
$N$-particle energy eigenstates $| 8 \rangle$ and $| 9 \rangle$ are almost 
exactly on resonance. Note that a Floquet state can be factorized 
according to  
\begin{eqnarray}
	& & 
	| u_j(t) \rangle\exp(-\ri\varepsilon_j t/\hbar) 
\nonumber \\	& = &	
	 | u_j(t) \re^{i m \omega t}\rangle
	 \exp(-\ri[\varepsilon_j + m\hbar\omega] t/\hbar)  
\label{eq:BZS}
\end{eqnarray}   
with an arbitrary positive or negative integer $m$, so that the Floquet
function $|u_j(t) \re^{i m \omega t}\rangle$ remains $T$-periodic, with
$T = 2\pi/\omega$. This means, loosely speaking, that ``the quasienergies are 
defined only up to an integer multiple of $\hbar\omega$.'' More precisely, 
the quasienergy of a Floquet state labeled by~$j$ has to be regarded as an 
infinite set of representatives $\varepsilon_j + m\hbar\omega$ spaced by 
$\hbar\omega$, implying that each Brillouin zone of the quasienergy spectrum 
of width $\hbar\omega$ contains precisely one representative of each state.

\begin{figure}[t]
\begin{center}
\includegraphics[width = 0.65\linewidth]{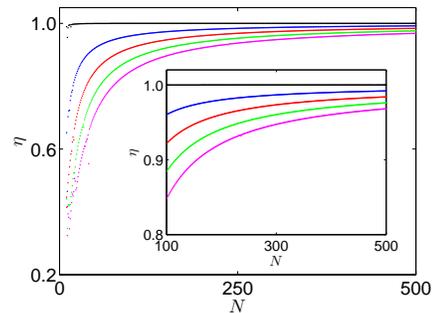}
\end{center}
\caption{(Color online) Degree of coherence~(\ref{eq:ETA}) for the 
	near-resonant Floquet states with Mathieu quantum numbers
	$k = 0, \ldots, 4$ (top to bottom) of the driven bosonic Josephson 
	junction~(\ref{eq:DJJ}) with $N\kappa/\Omega = 0.95$ kept fixed, 
	$\omega/\Omega = 1.62$, and $2\mu_1/\omega = 0.3$, vs.\ particle
	number~$N$.} 
\label{F_3}
\end{figure}

The Brillouin zone of quasienergies displayed in Fig.~\ref{F_2}~(a) features
a regular fan of almost equidistant lines, which can be explained analytically
by means of a standard resonance approximation~\cite{BermanZaslavsky77,
Holthaus95,GertjerenkenHolthaus14,GertjerenkenHolthaus14b}.
In the vicinity of the state $| r \rangle$ singled out by the condition
$\hbar\omega = E_{r+1} - E_r$, the dynamics of the driven $N$-particle system 
can be mapped to that of an effective quasiparticle, named ``floton'', which 
moves in a cosine potential well without external driving, such that the 
energies of this quasiparticle yield the quasienergies of the near-resonant 
Floquet states~\cite{GertjerenkenHolthaus14,GertjerenkenHolthaus14b}:   
\begin{equation} 
	\varepsilon_{k} = E_r + \frac{1}{8}E''_r\alpha_k(q) 
	\quad \mathrm{mod}\; \hbar\omega \; ,
\label{eq:MAP}
\end{equation}
where $E''_r$ denotes the formal (discrete) second derivative of the
unperturbed eigenvalues $E_j$ with respect to the state label~$j$, evaluated
at the resonant state $j = r$, and $\alpha_k(q)$ is a characteristic value
of the Mathieu equation. Using the notation of Ref.~\cite{AbramowitzStegun72},
one has $\alpha_k(q) = a_k(q)$ for quantum numbers $k = 0,2,4\ldots$ labeling
the even eigenstates of the floton quasiparticle, while $\alpha_k(q) = b_k(q)$
for $k = 1,3,5,\ldots\;$. The Mathieu parameter $q$ is proportional to the
driving amplitude,
\begin{equation}
	q = \frac{2}{E''_r/(\hbar\omega)} \frac{2\mu_1}{\omega}
	\langle r | a_1^{\dagger} a^{\phantom{\dagger}}_1 
	- a_2^{\dagger} a^{\phantom{\dagger}}_2 | r - 1 \rangle \; .
\end{equation}
The important feature here is the appearance of a new quantum number $k$: The
resonant state $| r \rangle$ turns into the floton ground state $k = 0$; the
neighboring states of the unperturbed junction~(\ref{eq:UDJ}) are transformed
into its excitations $k > 0$. In Fig.~\ref{F_3} we depict the degree of
coherence~(\ref{eq:ETA}) for the exact near-resonant Floquet states, computed
numerically, with floton quantum numbers $k = 0, \ldots, 4$. The similarity
to the previous Fig.~\ref{F_1} is striking: Indeed the ``resonant ground 
state'' $k = 0$ is an almost coherent state, in the sense that it corresponds
to an $N$-fold occupied, periodically time-dependent single-particle orbital.
Thus, here we encounter an example of Floquet engineering: The driving is not 
employed primarily to excite the system, but rather to create a new effective 
Hamiltonian~\cite{GoldmanDalibard14}, describing the floton quasiparticle, and 
providing a new ground state into which the actual particles can condense. 
This Floquet condensate constitutes a collective mode of response to the drive
which remains perfectly coherent in the course of time.

\section{Order parameter fluctuations}
  
\begin{figure}[t]
\begin{center}
\includegraphics[width = 0.65\linewidth]{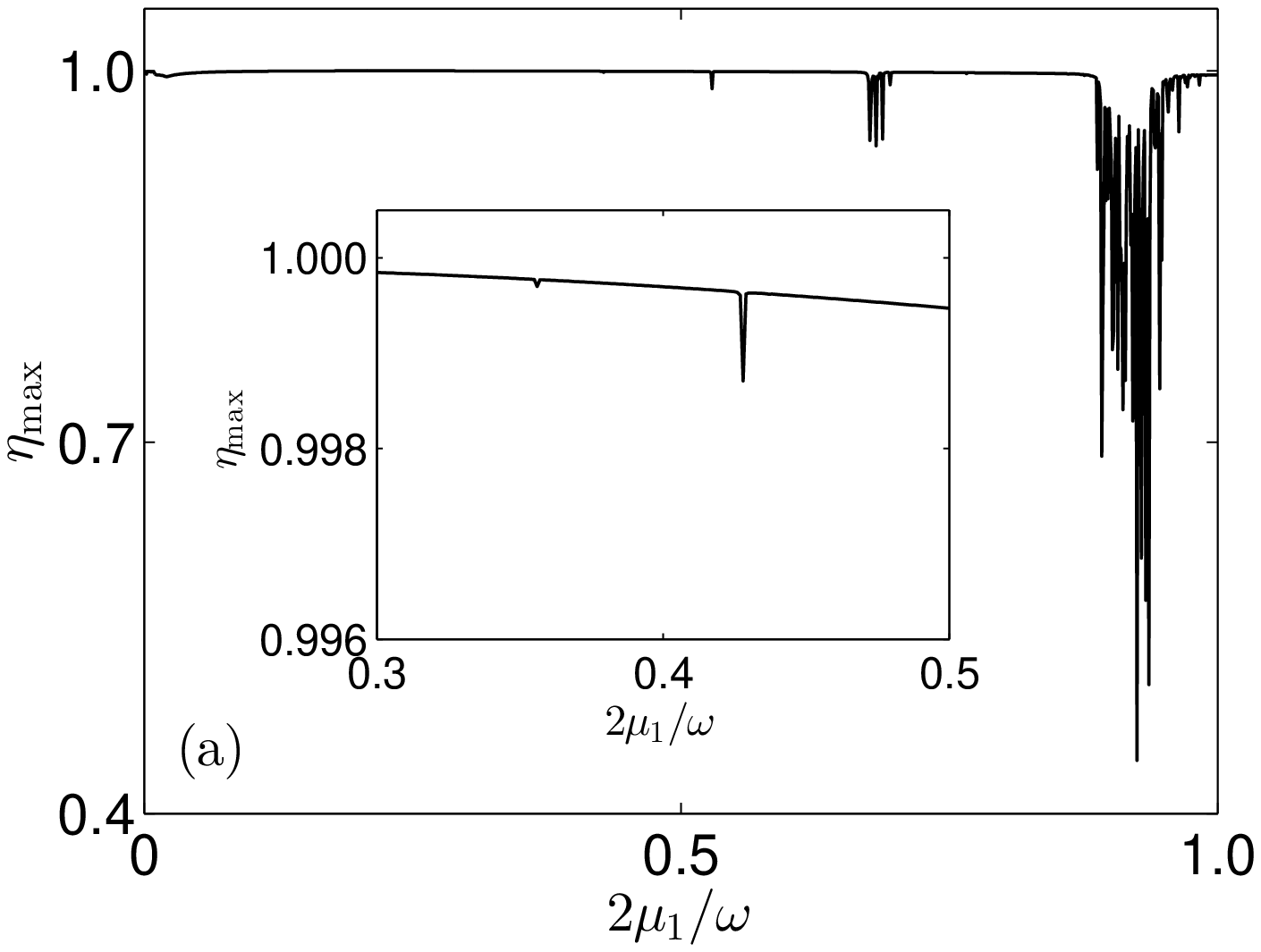}
\includegraphics[width = 0.65\linewidth]{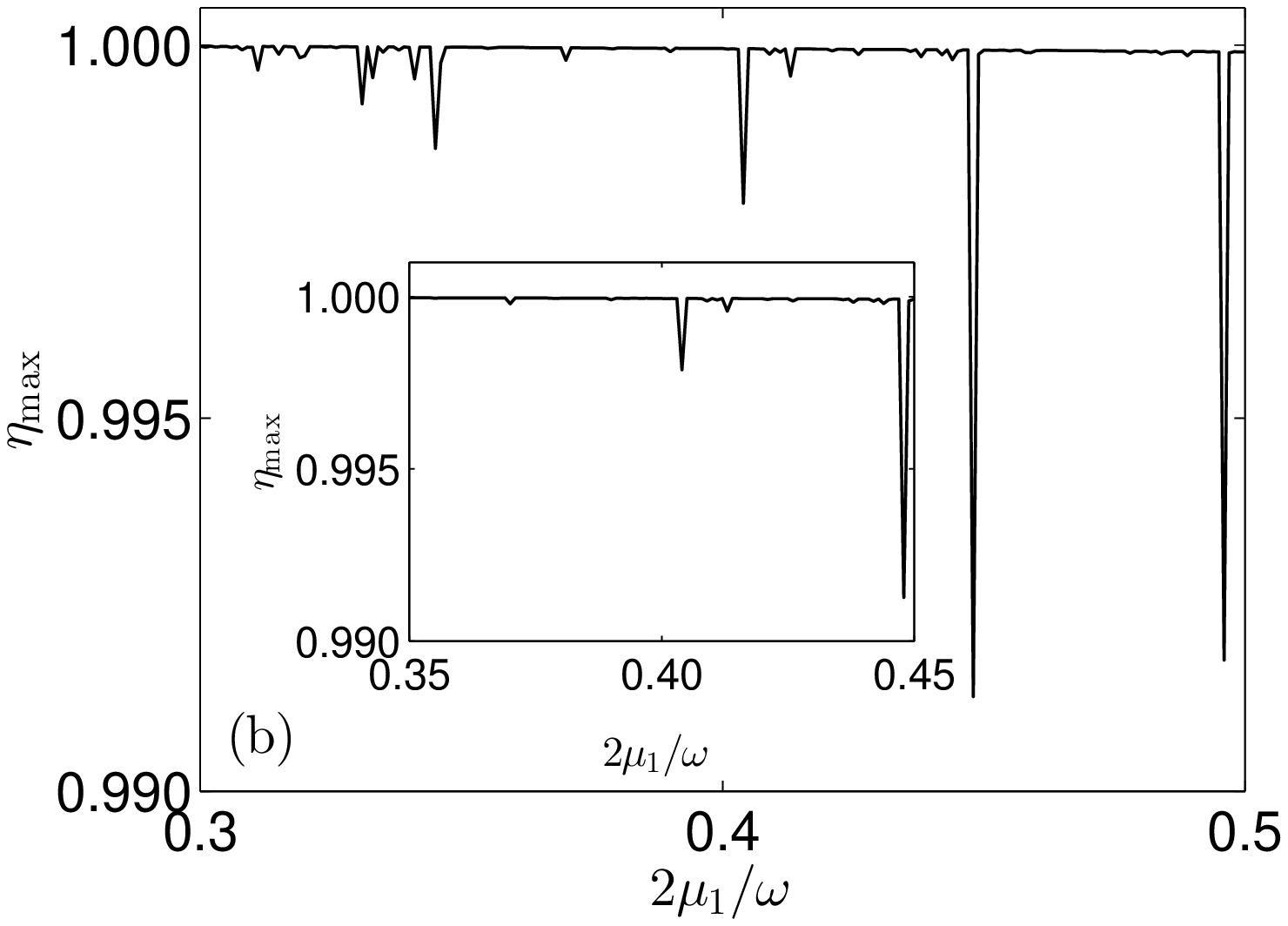}
\includegraphics[width = 0.65\linewidth]{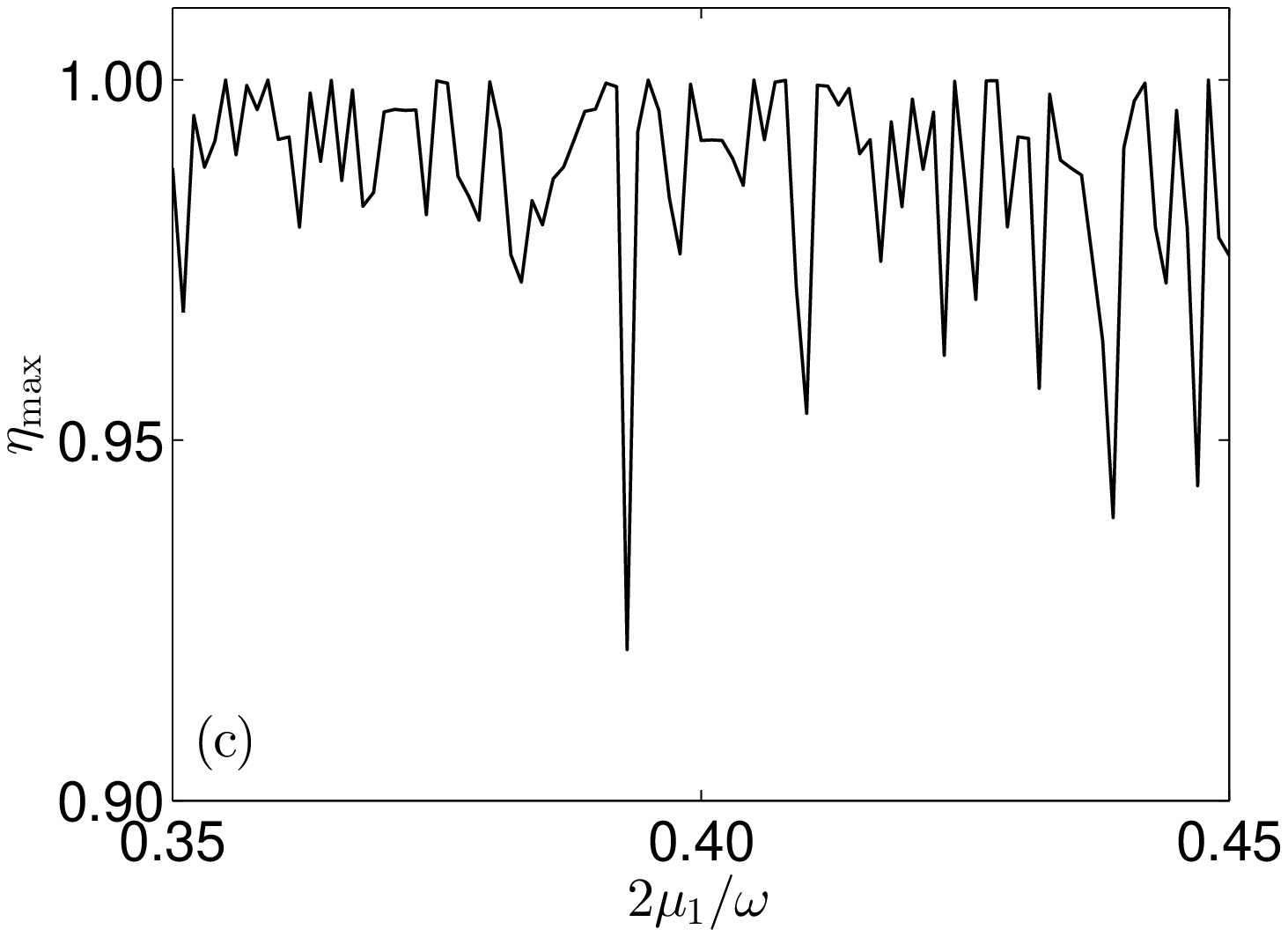}
\includegraphics[width = 0.65\linewidth]{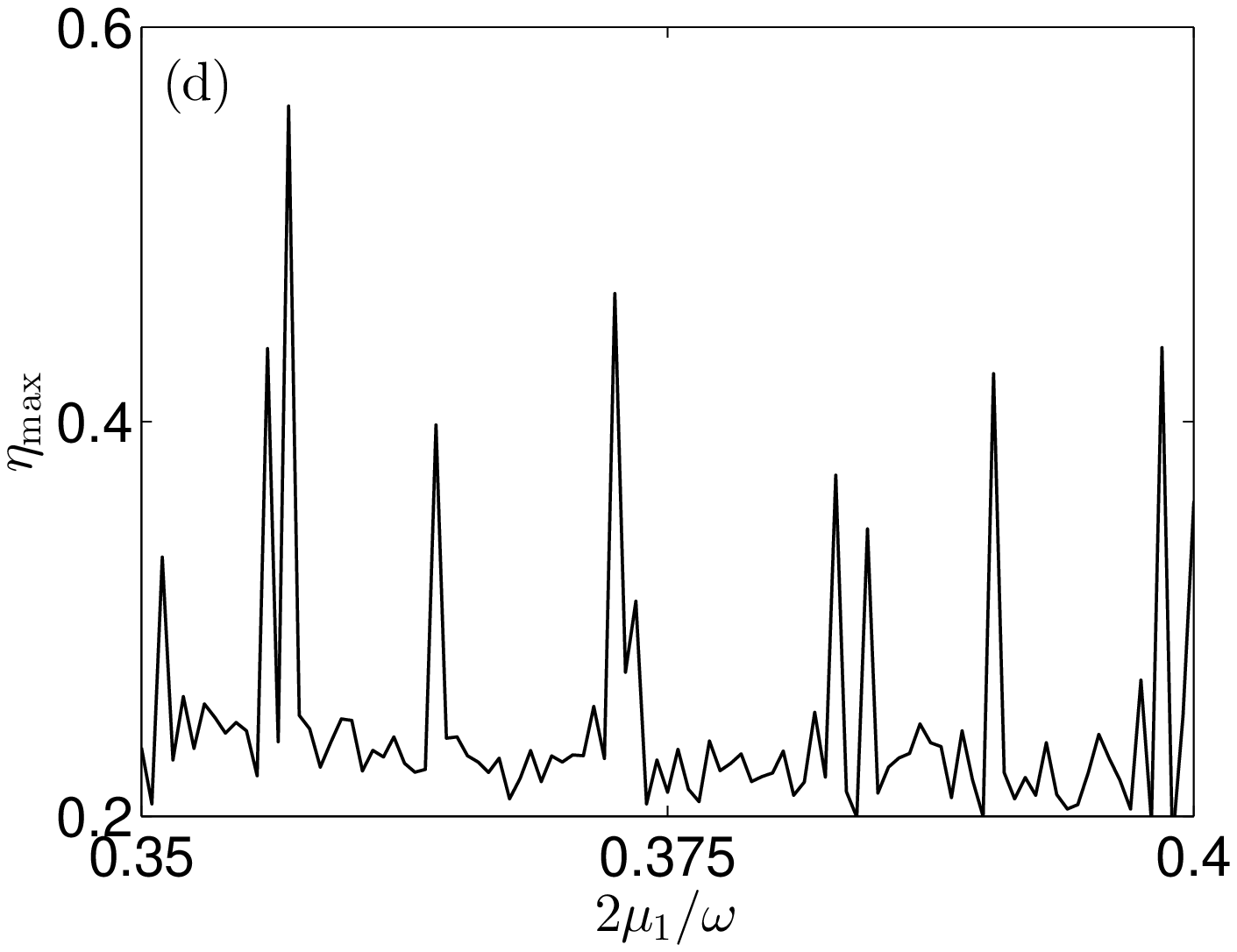}
\end{center}
\caption{Maximum degree of coherence~(\ref{eq:ETA}) of all $N$-particle
	Floquet states for $N\kappa/\Omega = 0.95$ and $\omega/\Omega = 1.62$.
	(a) $N = 100$; the inset delimits the interval of driving strengths 
	inspected in  the following panel. (b) $N = 500$; again the inset 
	marks the interval investigated in the following panel. (c) 
	$N = 1000$. (d) $N = 2000$. Observe the change of scale in comparison 
	to (c), and the shift of the baseline.}   
\label{F_4}
\end{figure}

However, there is a fundamental difference between such Floquet condensates 
and the customary, time-independent Bose-Einstein condensates which shows up 
if one tries to recover the mean field regime: In Fig.~\ref{F_4} we show the 
maximum degree of coherence $\eta_{\rm max}$, taken over all Floquet states 
of the driven Josephson junction~(\ref{eq:DJJ}) with $\omega/\Omega = 1.62$, 
vs.\ the scaled driving strength; again the interaction strength is adjusted
such that $N\kappa/\Omega = 0.95$. In panel~(a) we take $N = 100$: Here we 
observe extended intervals where $\eta_{\rm max} = 1$ with high accuracy, 
caused by the floton state $k = 0$, and large fluctuations occurring when 
$2\mu_1/\omega \approx 0.9$. The interval magnified in the inset is scanned 
again in panel~(b), but now with $N = 500$; here additional small fluctuations 
appear. Iterating this procedure, the interval framed in the inset of panel~(b) 
is evaluated in panel~(c) with $N = 1000$; here the fluctuations become more 
violent. In panel~(d), where $N = 2000$, even the baseline of the fluctuations 
is shifted downward. These results indicate that the size of a 
resonant Floquet condensate remains restricted to mesoscopically large 
particle numbers, while its order parameter would be destroyed for high~$N$ 
by large fluctuations.

The origin of these fluctuations is closely related to Eq.~(\ref{eq:BZS}), 
that is, to the Brillouin-zone structure of the quasienergy spectrum: Each 
zone contains $N+1$ quasienergy eigenvalues, as corresponding to the dimension 
of the junction's Hilbert space when there are $N$~Bose particles, so that the 
eigenvalue density is proportional to~$N$. On the other hand, eigenvalues 
falling into the same symmetry class are not allowed to cross. The quasienergy 
operator of the driven junction~(\ref{eq:DJJ}) remains invariant when the site 
labels are exchanged and simultaneously time is shifted by half a period; the 
Floquet functions therefore are even or odd under this generalized parity. 
Hence, neither ``odd'' nor ``even'' quasienergies may cross each other, which 
necessarily leads to a vast multitude of anticrossings when $N$ becomes large, 
each one indicating hybridization of the participating Floquet states. This
mechanism effectuates a degradation of the order parameter; each dip seen in 
panel~(b) can be traced to an isolated avoided quasienergy crossing. The 
Mathieu approximation~(\ref{eq:MAP}) locally reduces the driven $N$-particle 
system to an almost equivalent, integrable single-particle one, neglecting, 
in the sense of the rotating-wave approximation, fast-oscillating coupling 
terms~\cite{GertjerenkenHolthaus14,GertjerenkenHolthaus14b}. While for low 
driving amplitudes these couplings only produce anti\-crossings which are 
too small to detect on the scale of Fig.~\ref{F_2}~(a), their effect becomes 
stronger when $2\mu_1/\omega$ is increased. This eventually leads to a chaotic 
spectrum, as exemplified in Fig.~\ref{F_2}~(b). In the sequence shown in 
Fig.~\ref{F_4}, there are two opposing tendencies: On the one hand, the 
eigenvalue density increases by a factor of $20$ when enhancing $N$ from $100$ 
to $2000$; on the other, the interaction strength $\hbar\kappa$ is reduced by 
$1/20$. But evidently, this reduction is over-compensated by the growth of the 
particle number. While the individual anticrossings tend to become smaller upon
reducing $\hbar\kappa$, they proliferate and overlap upon increasing~$N$ to 
such an extent that the resulting multiple hybridizations forbid the formation 
of an order parameter: When the system becomes too complex, it does not possess
a simple mean field description. This absence of a proper mean field limit is 
closely related to the absence of an adiabatic limit in periodically driven 
quantum systems~\cite{HoneEtAl97}.

\section{Conclusions}
    
Since the appearance of resonances is a generic feature of driven nonlinear 
quantum systems, we anticipate that the findings reported in this work are 
not restricted to our particular model~(\ref{eq:DJJ}). Thus, we may summarize 
our main results as follows: {\em (i)\/} Resonantly driven Bose gases allow 
the formation of nonequilibrium Bose-Einstein condensates, with the 
resonance-induced effective ground state corresponding to a mesoscopically
occupied, periodically time-dependent single-particle orbital; {\em (ii)\/}
the coherence of such condensates is destroyed when the particle number becomes 
large, a mean field limit cannot be reached. This non-existence of a mean field
limit should be detectable through large fluctuations of the system's coherence
in a series of measurements in which the particle number varies slightly from 
shot to shot.

\begin{acknowledgments}    
We acknowledge support from the Deutsche Forschungsgemeinschaft (DFG) through 
grant No.\ HO 1771/6-2. The computations were performed on the HPC cluster 
HERO, located at the University of Oldenburg and funded by the DFG through
its Major Research Instrumentation Programme (INST 184/108-1 FUGG), and by 
the Ministry of Science and Culture (MWK) of the Lower Saxony State.
\end{acknowledgments}


\begin{thebibliography}{99}

\bibitem{LandauLifshitz75} L. D. Landau and E. M. Lifshitz,
	{\em Statistical Physics: Part 1\/}
	(Third edition, Butterworth-Heinemann, Oxford, 1975).

\bibitem{Huang87} K. Huang,
	{\em Statistical Mechanics\/}
	(Second edition, Wiley, New York, 1987). 

\bibitem{PathriaBeale11} R. K. Pathria and P. D. Beale,
	{\em Statistical Mechanics\/}
	(Third edition, Butterworth-Heinemann, Oxford, 2011). 
	
\bibitem{VorbergEtAl13} D. Vorberg, W. Wustmann, R. Ketzmerick, and A. Eckardt,
	Phys. Rev. Lett. {\bf 111}, 240405 (2013).	

\bibitem{PenroseOnsager56} O. Penrose and L. Onsager,
	Phys. Rev. {\bf 104}, 576 (1956). 
	
\bibitem{Leggett01} A. J. Leggett,
	Rev. Mod. Phys. {\bf 73}, 307 (2001).	
	
\bibitem{Shirley65} J. H. Shirley	
	Phys. Rev. {\bf 138}, B979 (1965).	
	
\bibitem{Zeldovich67} Ya. B. Zel'dovich, 
	Sov. Phys. JETP {\bf 24}, 1006 (1967)
	[Zh. Eksp. Teor. Fiz. {\bf 51}, 1492 (1966)].	
		
\bibitem{Sambe73} H. Sambe,
	Phys. Rev. A {\bf 7}, 2203 (1973).

\bibitem{FainshteinEtAl78} A. G. Fainshtein, N. L. Manakov, and L. P. Rapoport,
	J. Phys. B {\bf 11}, 2561 (1978).
	
\bibitem{LignierEtAl07} H. Lignier, C. Sias, D. Ciampini, Y. Singh, 
	A. Zenesini, O. Morsch, and E. Arimondo,
	Phys. Rev. Lett. {\bf 99}, 220403 (2007).
	
\bibitem{EckardtEtAl09} A. Eckardt, M. Holthaus, H. Lignier, A. Zenesini,
	D. Ciampini, O. Morsch, and E. Arimondo,
	Phys. Rev. A {\bf 79}, 013611 (2009).
	
\bibitem{ArimondoEtAl12}	
	E. Arimondo, D. Ciampini, A. Eckardt, M. Holthaus, and O. Morsch,
	Adv. At. Mol. Opt. Phys. {\bf 61}, 515 (2012).
			
\bibitem{ZenesiniEtAl09} A. Zenesini, H. Lignier, D. Ciampini, O. Morsch,
	and E. Arimondo,
	Phys. Rev. Lett. {\bf 102}, 100403 (2009).
	
\bibitem{AlbertiEtAl09} A. Alberti, V. V. Ivanov, G. M. Tino, and G. Ferrari,
	Nature Physics {\bf 5}, 547 (2009).
	
\bibitem{HallerEtAl10} E. Haller, R. Hart, M. J. Mark, J. G. Danzl, 
	L. Reich\-s\"ollner, and H.-C. N\"agerl,
	Phys. Rev. Lett. {\bf 104}, 200403 (2010).		

\bibitem{StruckEtAl11} J. Struck, C. \"Olschl\"ager, R. Le Targat, 
	P. Soltan-Panahi, A. Eckardt, M. Lewenstein, P. Windpassinger, and
	K. Sengstock, 
	Science {\bf 333}, 996 (2011). 
	
\bibitem{ChenEtAl11} Y.-A.\ Chen, S. Nascimb\`ene, M. Aidelsburger, M. Atala, 
	S. Trotzky, and I. Bloch,
	Phys. Rev. Lett. {\bf 107}, 210405 (2011).
	
\bibitem{StruckEtAl12} J. Struck, C. \"Olschl\"ager, M. Weinberg, P. Hauke,
	J. Simonet, A. Eckardt, M. Lewenstein, K. Sengstock,
	and P. Windpassinger,
	Phys. Rev. Lett. {\bf 108}, 225304 (2012).
		
\bibitem{StruckEtAl13} J. Struck, M. Weinberg, C. \"Olschl\"ager,
	P. Windpassinger, J. Simonet, K. Sengstock, R. H\"oppner, P. Hauke,
	A. Eckardt, M. Lewenstein, and L. Mathey,
	Nature Physics {\bf 9}, 738 (2013).
	
\bibitem{ParkerEtAl13} C. V. Parker, L.-C. Ha, and C. Chin,
	Nature Physics {\bf 9}, 769 (2013).			 		
		 		
\bibitem{GatiMKO07} R. Gati and M. K. Oberthaler,
	J. Phys. B {\bf 40}, R61 (2007).		
				
\bibitem{LipkinEtAl65} H. J. Lipkin, N. Meshkov, and A. J. Glick,
	Nuc. Phys. {\bf 62}, 188 (1965).
	
\bibitem{MilburnEtAl97} G. J. Milburn, J. Corney, E. M. Wright, 
	and D. F. Walls,
	Phys. Rev. A {\bf 55}, 4318 (1997).

\bibitem{ParkinsWalls98} A. S. Parkins and D. F. Walls,
	Phys. Rep. {\bf 303}, 1 (1998).
	
\bibitem{HolthausStenholm01}
	M. Holthaus and S. Stenholm,
	Eur. Phys. J. B {\bf 20}, 451 (2001).
			
\bibitem{JinasunderaEtAl06}
	T. Jinasundera, C. Weiss, and M. Holthaus,
	Chem. Phys. {\bf 322}, 118 (2006).	

\bibitem{MahmudEtAl05}
	K. W. Mahmud, H. Perry, and W. P. Reinhardt, 
	Phys. Rev. A {\bf 71}, 023615 (2005).
	
\bibitem{BoukobzaEtAl10} E. Boukobza, M. G. Moore, D. Cohen, and A. Vardi,
	Phys. Rev. Lett. {\bf 104}, 240402 (2010).
				
\bibitem{LiebEtAl00} E. H. Lieb, R. Seiringer, and J. Yngvason,
	Phys. Rev. A {\bf 61}, 043602 (2000).		
		
\bibitem{BermanZaslavsky77}
	G. P. Berman and G. M. Zaslavsky,
	Phys. Lett. A {\bf 61}, 295 (1977).	
	
\bibitem{Holthaus95}
	M. Holthaus, 
	Chaos, Solitons \& Fractals {\bf 5}, 1143 (1995).
	
\bibitem{GertjerenkenHolthaus14}
	B. Gertjerenken and M. Holthaus, 	
	New J.\ Phys.\ {\bf 16}, 093009 (2014).	
	
\bibitem{GertjerenkenHolthaus14b}
	B. Gertjerenken and M. Holthaus,
	preprint (submitted to Phys. Rev. A). 	

\bibitem{AbramowitzStegun72}
	M. Abramowitz and I. A. Stegun (Eds.),  
	{\em Handbook of Mathematical Functions\/}
	(Dover, New York, 1972), ch.~20.
	
\bibitem{GoldmanDalibard14}
	N. Goldman and J. Dalibard,
	Phys. Rev. X {\bf 4}, 031027 (2014).	
	
\bibitem{HoneEtAl97} D. W. Hone, R. Ketzmerick, and W. Kohn,
	Phys. Rev. A {\bf 56}, 4045 (1997).	
		
\end{thebibliography}
\end{document}